\documentstyle[12pt]{article}
\def\be {\begin{equation}}
\def\ee {\end{equation}}
\newtheorem{mythname}{Theorem}
\newtheorem{coro}{Corollary}
\newtheorem{rem}{Remark}
\begin{document}
\title{Poisson Structures for Dispersionless Integrable Systems and
       Associated W-Algebras}
\author{Yi Cheng \\
     {\small Department of Mathematics, University of Science and}\\
     {\small Technology of China, Hefei, Anhui, 230026, P.R.China}\\
     {\small E-mail: chengy@hpe25.ustc.edu.cn}\\
        Zhifeng Li \\
     {\small Nonlinear Science Center, University of Science and}\\
     {\small Technology of China, Hefei, Anhui, 230026, P.R.China}\\
     {\small E-mail: zfli@sunlx06.nsc.ustc.edu.cn}}
\maketitle
\begin{center}
\begin{minipage}{120mm}
\vskip 0.1in
\begin{center}{\bf Abstract}\end{center}
{In analogy to the KP theory, the second Poisson structure for the dispersionless
KP hierarchy can be defined on the space of commutative
pseudodifferential operators $L=p^n+\sum_{j=-\infty}^{n-1}u_j p^j$. The
reduction of the Poisson structure to the symplectic submanifold $u_{n-1}=0$ gives rise to the
w-algebras. In this paper, we discuss properties of this Poisson structure,
its Miura transformation and reductions. We are particularly interested in the
following two cases: a) $L$ is pure polynomial in $p$ with multiple roots and
b) $L$ has multiple poles at finite distance. The w-algebra corresponding to
the case a) is defined as ${\rm w}_ {[m_1,m_2,\cdots ,m_r]}$, where $m_i$ means the
multiplicity of roots and to the case b) is defined by
${\rm w}(n,[m_1,m_2,\cdots,m_r])$ where $m_i$ is the multiplicity of poles.
We prove that ${\rm w}(n, [m_1,m_2,\cdots, m_r])$-algebra is isomorphic
via a transformation to ${\rm w}_{[m_1,m_2,\cdots,m_r]} \bigoplus $${\rm w}_{n+m}
\bigoplus {\rm U(1)}$ with $m=\sum m_i$. We also give the explicit free fields
representations for these w-algebras }
\end{minipage}
\end{center}
\newpage
It is well-known that the second Poisson structures of integrable hierarchies
give rise to the classical realizations of conformal W-algebras. The Miura
transformation that transforms
the second Poisson structures of systems to vastly simpler ones provides
the free fields representations of correspondent W-algebras. (see[1-3] for
reviews)
\par
   The purpose of this paper is to discuss some features  dealing with the (second)
Poisson structure for the dispersionless Kadomtsev-Petviashvili (dKP) hierarchy,
its reductions and associated w-algebras. We are particularly interested in
the reduced case that the Lax pair of the hierarchy have either multiple roots or
multiple poles at finite distance. \par
   The dKP hierarchy is known as the dispersionless limit of the (dispersionful)
KP hierarchy. It is written in the following Lax form [4,5,6]
\be
\label{e1}
\frac{\partial L}{\partial t_n}=[[(L^n)_{+}, L]],
\ee
where $L$ is the commutative pseudodifferential
operators (c$\Psi $DOs)
\be
\label{e2}
L=p+ \sum\limits_{j=1}^{\infty}u_{j}p^{-j}
\ee
and the subscript "+" is defined as usual to take the polynomial part of a
c$\Psi $DO. The double bracket in (1) is defined by
\be
\label{e3}
[[A,B]]=\frac{\partial A}{\partial p}\frac{\partial B}{\partial x}-\frac{\partial
 A}{\partial x}\frac{\partial B}{\partial p},
\ee
for any two c$\Psi $DOs $A$ and $B$ with coefficients being smooth functions
of $x$. Similar to the KP hierarchy, the dKP hierarchy admits some reductions.
If $L^n$ is restricted to be polynomial in $p$ for a fixed integer $n$, then
dKP is reduced to the dispersionless Gelfand-Dickey hierarchy [6]. The Zakharov
reduction is represented by the restriction that $L^n$ has multiple poles at
finite distance [7]
\be
\label{e4}
L^n=p^n+\sum\limits_{j=0}^{n-2}u_j p^j +\sum\limits_{i=1}^r\sum\limits_{j=1}^{m_i}
{\frac{a_{i,j}}{(p+p_i)^j}}
\ee
and the associated system is called the extended Benney hierarchy. \par
The Poisson structures for dKP are built first on the space $M _{n}$ of
\par
\be
\label{e5}
L=p^n+\sum\limits_{j=-\infty}^{n-1}u_j p^j
\ee
and then on the submanifold $u_{n-1}=0$ by constraint. They can be formulated either
by taking dispersionless limit of the Poisson structures for the KP hierarchy [8]
or by a dispersionless analogue of the Adler-Gelfand-Dickey construction [4,5].
In particular the second Poisson structure on the space $M_n$ is given by
\par
\begin{eqnarray}
\{\tilde{f},\tilde{g}\}^{(n)}&=&\int{\rm res} ([[(\frac{\delta{f}}{\delta{L}}L)_+ ,L]]
\frac{\delta{g}}{\delta{L}}-[[\frac{\delta{f}}{\delta{L}},L]]_+ L\frac{\delta{g}}
{\delta{L}})dx \nonumber  \\
&=&\int{\rm res} ([[(\frac{\delta{f}}{\delta{L}}L)_+ ,L]]\frac{\delta{g}}{\delta{L}}
+[[(\frac{\delta{g}}{\delta{L}}L)_- ,L]]\frac{\delta{f}}{\delta{L}})dx,
\end{eqnarray}
where $L$ is in the form of (5), $\tilde{f}=\int f(u)dx$ with $f(u)=
f(u_{n-1},u_{n-2},\cdots,)$ being polynomial in $(u_{n-1},u_{n-2},\cdots)$. The variation
${\delta{f}}/{\delta{L}}$ is defined by
\be
\label{e7}
\frac{\delta{f}}{\delta{L}}=\sum\limits_{j=-\infty}^{n-1}\frac{\partial{f}}{\partial u_j}
p^{-j-1}
\ee
and similarly for $\tilde{g}$ and ${\delta g}/{\delta L}$. The second equality
in (6) was obtained by using
\be
\label{e8}
\int{\rm res} ([[A,B]]C)dx=\int{\rm res} ([[B,C]]A)dx
\ee
for arbitrary $A$, $B$ and $C$. If (6) is restricted to $u_{n-1}=0$,
then the following condition [4,5,8]
\be
\label{e9}
{\rm res}[[\frac{\delta{f}}{\delta{L}}, L]]=0
\ee
must to be taken into account. From (6), the Poisson brackets among fields \newline
$(u_{n-1},u_{n-2},\cdots)$ are expressed by
\be
\label{e10}
\{u_i(x), u_j(y)\}^{(n)}=J_{ij}^{(n)}(u)\delta (x-y),  i,j \leq n-1
\ee
which provides the ${\rm w}_{{\rm dKP}}^{(n)}$-algebras. The constraint $u_{n-1}=0$ is the
second class one and the reduced brackets
\be
\label{e11}
\{u_i(x), u_j(y)\}_D^{(n)}=\tilde{J}_{ij}^{(n)}(u)\delta (x-y),  \ \ i,j \leq n-2
\ee
represent the
$\hat{\rm w}_{\infty}^{(n)}$-algebra, where
\be
\label{e12}
\tilde{J}_{ij}^{(n)}=J_{ij}^{(n)}-J_{i,n-1}^{(n)} (J_{n-1,n-1}^{(n)})^{-1} J_{n-1,j}^{(n)}.
\ee
We will show both ${\rm w}_{{\rm dKP}}^{(n)}$  and $\hat{\rm w}_{\infty}^{(n)}$
are independent on $n$.
\par
In this paper, analogous to our previous construction for the KP hierarchy
[9,10], We give a general Miura transformation by expressing $L \in M_n$
into a factorization from $L=L_1 L_2$ where $L_j \in M_{n_{j}}$ and
$n_1 +n_2=n$. As a result the second Poisson brackets (6) is decomposed
into a summation of two brackets associated with $L_1$ and $L_2$
respectively. It is then no difficulty to extend the above factorization to
a rational from $L=L_1L_2^{-1}$, since in this case, we may think that $L_1$
is factorized as $L_1=L L_2$. The rational factorization will give rise to
the Miura transformation for the extended Benney hierarchy, which also leads
to a decomposition of the associated second Poisson structure. In other words,
we may decompose the correspondent w-algebra to a direct sum of others.
\par
we emphasize that although in the generic case, the dKP and associated w-algebra
can be obtained from the counterpart for the KP by taking dispersionless limit,
as far as we know, there is no similar result for the systems, such as the extended Benney hierarchy,
that the associated c$\Psi $DO $L$  has multiple roots or multiple poles.
\par
\vskip 0.3cm
Let $L$ in (5) is factorized by
\be
\label{e13}
L=L_1 L_2,
\ee
where $L_j=p^{n_j}+v_{j,n_j -1}p^{n_j -1}+\cdots, \ j=1,2$ are two c$\Psi $DOs
and $n_1 +n_2 =n$. Compare the same powers of $p$ in both sides of (13), we obtain
the Miura transformation
\begin{eqnarray}
\label{e14}
&u_{n-1}=v_{1,n_1 -1}+ v_{2,n_2-1}, \nonumber \\
&u_j=\sum\limits_{j-n_{1}+1 \leq k \leq n_1-1} v_{1,k}v_{2,j-k}.
\end{eqnarray}
\par
\begin{mythname}
The factorization (13) leads to the following decomposition of the second Poisson
bracket (6) associated with $L$ in (5)
\be
\label{e15}
\{\tilde{f}, \tilde{g}\}^{(n)}=\{\tilde{f}, \tilde{g}\}^{(n_1)}+\{\tilde{f}, \tilde{g}\}^{(n_2)},
\ee
and the constraint conditon $u_{n-1}=v_{1,n_1 -1} +v_{2,n_{2}-1}=0$ is equivalent
to
\be
\label{e16}
{\rm res}[[\frac{\delta{f}}{\delta{L}}, L]]=
{\rm res}[[\frac{\delta{f}}{\delta{L_1}}, L_1]]
 +{\rm res}[[\frac{\delta{f}}{\delta{L_2}},L_2]]=0.
\ee
\end{mythname}
\par
Proof: First we can express ${\delta{f}}/{\delta{L_j}}$ in terms of
${\delta{f}}/{\delta{L}}$
\be
\label{e17}
\frac{\delta{f}}{\delta{L_1}}=\frac{\delta{L}}{\delta{L}}L_2, \ \
\frac{\delta{f}}{\delta{L_2}}=\frac{\delta{f}}{\delta{L}}L_1.
\ee
They can be derived by using the Miura transformation (14)
\begin{eqnarray*}
\label{e*}
\frac{\partial f}{\partial v_{1k}}=\sum\limits_j \frac{\partial f}{\partial u_j}
 \frac{\partial u_j}{\partial v_{1k}}=\sum\limits_j 
 \frac{\partial f}{\partial u_j}v_{2,j-k}, \nonumber
\end{eqnarray*}
so
\be
\label{e18}
\frac{\delta f}{\delta L_1}=\sum\limits_k \frac{\partial f}{\partial v_{1k}}p^{-k-1}
=\sum\limits_{k}\sum\limits_{j}\frac{\partial f}{\partial u_j}v_{2,j-k}p^{-k-1}
=\frac{\delta f}{\delta L}L_2,
\ee
and similarly for the second expression in (17). Then by using the Leibeniz rule
\be
\label{e19}
[[A, BC]]=[[A, B]]C+[[A, C]]B
\ee
of the double square bracket (3), we have
\begin{eqnarray*}
&&[[(\frac{\delta f}{\delta L_1}L_1)_{\pm}, L_1]]\frac{\delta g}{\delta{L_1}}
+[[(\frac{\delta f}{\delta L}L)_{\pm}, L_2]]\frac{\delta g}{\delta L_2}\\
&=&([[(\frac{\delta f}{\delta L_1})_{\pm}, L_1]]L_2+[[(\frac{\delta f}{\delta L}L)_{\pm}
, L_2]]L_1)\frac{\delta g}{\delta L}\\
&=&[[(\frac{\delta f}{\delta L}L)_{\pm}, L]]\frac{\delta g}{\delta L}
\end{eqnarray*}
which immediately implies (15). The equation in (16) can also be derived by using
(17), i.e.
\be
\label{e20}
[[\frac{\delta f}{\delta L_1}, L_1]]+[[\frac{\delta f}{\delta L_2}, L_2]]
=[[\frac{\delta f}{\delta L}L_2, L_1]]+[[\frac{\delta f}{\delta L}L_1, L_2]]
=[[\frac{\delta f}{\delta L}, L]].
\ee
Thus we complete the proof.
\begin{coro}
If L is factorized in the following form
\be
\label{e21}
L=L_1 L_2\cdots L_r,
\ee
with $L_j$ being c$\Psi $DOs of order $n_j$, then
\be
\label{e22}
\{\tilde{f}, \tilde{g}\}^{(n)}=\sum\limits_{j=1}^r \{\tilde{f}, \tilde{g}\}^{(n_j)}
\ee
and the constraint condition $u_{n-1}=0$ is given by
\be
\label{e23}
{\rm res}[[\frac{\delta{f}}{\delta L}, L]]={\rm res}\sum\limits_{j=1}^{r}[[\frac{\delta f}{\delta L_j}, L_j]]=0.
\ee      
\end{coro}
\begin{coro}
If $L$ is factorized in the rational form
\be
\label{e24}
L=L_{1}L_{2}^{-1},
\ee
where $L_1$ and $L_2$ are $(n+m)^{th}-$order and $m^{th}-$order polynomials respectively
\begin{eqnarray}
\label{e25}
L_1&=&p^{n+m}+v_{n+m-1}p^{n+m-1}+\cdots +v_0; \nonumber \\
L_2&=&p^m+w_{m-1}p^{m-1}+\cdots + w_0,
\end{eqnarray}
then we have
\be
\label{e26}
\{\tilde{f}, \tilde{g}\}^{(n)}=\{\tilde{f}, \tilde{g}\}^{(n+m)}-\{\tilde{f}, \tilde{g}\}^{m}.
\ee
\end{coro}
\begin{mythname}
If $L$ is the $k^{th}$ power of the $\Psi $DO $L_1$ of order $l$
\be
\label{e27}
L=L_{1}^k,
\ee
then we have
\be
\label{e28}
\{\tilde{f}, \tilde{g}\}^{(kl)}=\frac{1}{k} \{\tilde{f}, \tilde{g}\}^{(l)}.
\ee
\end{mythname}
\par
Proof: Similar to the proof of Theorem 1, we first have
\be
\label{e29}
\frac{\delta f}{\delta L_1}=k\frac{\delta f}{\delta L}L_1^{k-1}
\ee
By substituting this expression into the right hand side of (26), we can derive (28).
\begin{rem}
Theorem 1 is the dispersionless analogue of our previous result for
the KP hierarchy [9,10] but the proof is simplified. The rational factorization and
their resulted decomposition formula of the second Poisson structure for the KP
hierarchy was particularly derived in [11], which can also be, however, considered as
the consequence of our previous result in [9,10].
\end{rem} 
\begin{rem}
If we choose $L_1=p+v_0+v_1 p^{-1}+\cdots$, then (26)
implies that the second Poisson structure associated with $L_1^k$ for any integer
$k$ is proportional to that associated with $L_1$. In other words, ${\rm w}_{dKP}^{(k)}$
-algebra is essentially independent on the value of $k$ [12]. There is no such
an analogue for the KP hierarchy and ${\rm W}_{KP}^{(k)}$-algebra, since we known that in the KP
theory, the second Poisson structrues associated with $L_1^k$ (here $L_1$ is
the non-commutative $\Psi $DOs) are not compatible with different
values of $k$ [13].
\end{rem}
In the following, we give some applications of the above general theory.
For the simplicity, we use notation $\{ \ , \  \}$ instead of the Poisson
br
acket $\{ \ , \ \}^{(n)}$ associated with generic c$\Psi $DO $L$. \par
\par
We first consider the Poisson structure and associated w-algebra on the space
of
\be
\label{e30}
L=p^m+\sum\limits_{j=1}^{m-1}u_{j}p^{j}=\prod\limits_{j=1}^{r}(p+p_j)^{m_j},
\ee
where
\be
\label{e31}
m=\sum\limits_{j=1}^{r}m_j,
\ee
namely we assume that $L$ is a polynomial with multiple roots. Notice that the Poisson
structure (6) for $L=p+p_j$ is simply $\partial _{x}$, thus by applying Corollary
2 and Theorem 2, the Poisson brackets among $p_j$ are
\be
\label{e32}
\{p_{i}(x), p_{j}(y)\}=\frac{1}{m_i}\delta _{ij} \delta ^{'}(x-y).
\ee
The constraint $u_{m-1}=\sum\limits_{j=1}^{r}m_j p_j =0$ is the second type and
the reduced Poisson brackets are
\be
\label{e33}
\{p_{i}(x), p_{j}(y)\}=(\frac{1}{m_i}\delta _{ij}-\frac{1}{m}) \delta ^{'}(x-y).
\ee
Since (30) provides expressions of $u_j$ in terms of $p_j$, therefore from (32)
one may derive the Poisson brackets among $u_{m-2},\cdots, u_0$. which is defiend
as ${\rm w}_{[m_1,\cdots,m_r]}$-algebra. If $r=m$ and all $m_j=1$, then we recover
${\rm w}_{m}$-algebra [12], i.e. ${\rm w}_{[1,\cdots,1]}={\rm w}_m$
\par
To present the free fields realization of ${\rm w}_{[m_1,\cdots,m_r]}$-algebra, we
introduce $r-1$ free fields
\be
\label{e34}
{\bf \phi}=(\varphi _{1},\cdots, \varphi _{r-1}),
\ee
with Poisson bracket
\be
\label{e35}
\{\varphi _{i}^{'}(x), \varphi _{i}^{'}(y)\}_{D}=\delta _{ij}\delta ^{'}(x-y),
\ee
and an overcomplete set of vectors ${\bf h}_j, \ j=1,2,\cdots,r$ in $(r-1)$-dimensional
Euclidean space with
\be
\label{e36}
\sum\limits_{j=1}^{r}m_{j}{\bf h}_j=0,  \ \ {\bf h}_i\cdot {\bf h}_j=(\frac{1}{m_i}
\delta_{ij}-\frac{1}{m}).
\ee
Such vectors can be written exactly, they are
\begin{eqnarray}
\label{37}
{\bf h}_1&=&((\frac{m_2}{m_1(m_1+m_2)})^{\frac{1}{2}},(\frac{m_3}
{(m_1+m_2)(m_1+m_2+m_3)})^{\frac{1}{2}},
\cdots, (\frac{m_r}{(\sum\limits_{i=1}^{r-1}m_i)
(\sum\limits_{i=1}^{r}m_i)})^{\frac{1}{2}}); \nonumber \\
{\bf h}_2&=&(-(\frac{m_1}{m_2(m_1+m_2)})^{\frac{1}{2}},
(\frac{m_3}{(m_1+m_2)(m_1+m_2+m_3)})^{\frac{1}{2}},
\cdots, (\frac{m_r}{(\sum\limits_{i=1}^{r-1}m_i)(\sum\limits_{i=1}
^{r}m_i)})^{\frac{1}{2}});  \nonumber \\
{\bf h}_3&=&(0, \ -(\frac{m_1+m_2}{m_3(m_1+m_2)})^{\frac{1}{2}},(\frac{m_4}
{(\sum\limits_{i=1}^{3}m_i)(\sum\limits_{i=1}^{4}m_i)})^{\frac{1}{2}},
\cdots, (\frac{m_r}{(\sum\limits_{i=1}^{r-1}m_i)(
\sum\limits_{i=1}^{r}m_i)})^{\frac{1}{2}}); \\
&&\cdots \ \cdots \nonumber\\
{\bf h}_j&=&(0, \ \cdots, \ 0, \ -(\frac{\sum\limits_{i=1}^{j-1}m_i}{m_j\sum\limits_{i=1}
^{j}m_i})^{\frac{1}{2}},(\frac{m_{j+1}}{(\sum\limits_{i=1}^{j}m_i)
(\sum\limits_{i=1}^{j+1}m_j)})^{\frac{1}{2}},
\cdots, (\frac{m_r}{(\sum\limits_{i=1}^{r-1}m_i)(\sum\limits_{i=1}
^{r}m_i)})^{\frac{1}{2}}); \nonumber \\
&&\cdots \ \cdots  \nonumber \\
{\bf h}_r&=&(0, \ \cdots,\ 0, \ -(\frac{\sum\limits_{i=1}^{r-1}m_i}{(m_r
\sum\limits_{i=1}^{r}m_i)})^{\frac{1}{2}}). \nonumber
\end{eqnarray}
Then using the identification
\be
\label{e38}
p^m+\sum\limits_{j=1}^{m-2}u_{j}p^{j}=\prod\limits_{j=1}^{r}(p+{\bf h}_j\cdot
{\bf \phi}^{'})^{m_j},
\ee
we obtain the free fields realization of the ${\rm w}_{[m_1,\cdots,m_r]}$ -algebra.
The particular expression $u_{m-2}$ is straightforward to derive from (38),
\be
\label{e39}
u_{m-2}=-\frac{1}{2}\sum\limits_{j=1}^{r}m_{j}({\bf h}_{j}\cdot {\bf \phi }^{'})^{2}.
\ee
It satisfies
\be
\label{e40}
\{u_{m-2}(x), u_{m-2}(y)\}_{D}=-(u_{m-2}(x)\partial + \partial u_{m-2}(x))\delta (x-y).
\ee
\par
Next we consider the Poisson structure on the space of
\be
\label{e41}
L=p^n+\sum\limits_{j=0}^{n-1}u_{j}p^j +\sum\limits_{i=1}^{r}\sum\limits_{j=1}^{m_i}
\frac{a_{ij}}{(p+p_i)^j},
\ee
and then take the reduction $u_{n-1}=0$, where
\be
\label{e42}
\sum\limits_{j=1}^rm_j =m,
\ee
and $p_j$ are distinct poles. Such type of $L$ can be written as
\be
\label{e43}
L(p)=L_{1}(p)L_{2}^{-1}(p),
\ee
with
\be
\label{e44}
L_{1}(p)=\prod\limits_{j=1}^{n+m}{(p+q_j)}; \ \
L_{2}(p)=\prod\limits_{j=1}^{r}(p+p_j)^{m_j},
\ee
here for the simplicity we assume all $q_j$ are distinct. By the application of
Corollary 2 and results associated with (29). we find
\begin{eqnarray}
\label{e45}
\{p_i, p_j\}&=&-\frac{1}{m_i}\delta _{ij}\delta^{'}; \nonumber \\
\{q_k, q_l\}&=&\delta _{ij}\delta^ {'};   \\
\{p_i, q_k\}&=&0, \nonumber 
1\leq i,j\leq r,  1\leq k,l\leq n+m. \nonumber
\end{eqnarray}
The constraint
\be
\label{e46}
u_{m-1}=\sum\limits_{k=1}^{n+m}q_k -\sum\limits_{j=1}^{r}m_j p_j=0
\ee
is the second type and the reduced brackets are given by
\begin{eqnarray}
\label{e47}
\{p_i, p_j\}_D&=&-(\frac{1}{m_i}+ \frac{1}{n})\delta^{'}; \nonumber\\
\{q_k, q_l\}_D&=&(\delta _{kl}- \frac{1}{n})\delta^{'}; \\
\{p_i, q_k\}_D&=&- \frac{1}{n}\delta^{'},      \nonumber
\end{eqnarray}
for $1\leq i,j\leq r, \ \ 1\leq k,l\leq n+m$.
We denote the associated w-algebra by ${\rm w}(n,[m_1,\cdots, m_r]) $.
\begin{mythname}
The ${\rm w}{(n,[m_1,\cdots, m_r])}$-algebra is isomorphic via a transformation to the
direct sum of a ${\rm w}_{[m_1,\cdots, m_r]}$-algebra, a ${\rm w}_{n+m}$-algebra and a
U(1) current algebra.
\end{mythname}
\par
Proof: Let
\par
\begin{eqnarray}
\label{e48}
J&=&\frac{2}{m(m+n)}\sum\limits_{i=1}^{r}m_i p_i =\frac{2}{m(m+n)}\sum\limits_{k=1}^{m+n}q_k;   \nonumber \\
\bar{p}_i&=&p_i -\frac{m+n}{2}J, \ \   1\leq i\leq r; \\
\bar{q}_k&=&q_k -\frac{m}{2}J,    \ \  1\leq k\leq m+n. \nonumber
\end{eqnarray}
One can check that
\be
\label{e49}
\sum\limits_{i=1}^{r}m_{i}\bar{p_i}=0,  \ \
\sum\limits_{k=1}^{m+n}\bar{q_k}=0.
\ee
\par
The Poisson brackets among these new fields are given by
\begin{eqnarray}
\label{e50}
\{J, J\}_D&=&-(\frac{4}{mn(n+m)})\delta^{'}; \nonumber\\
\{\bar{p}_i, \bar{p}_j\}_D&=&-(\delta _{ij}- \frac{1}{m})\delta^{'},
\ \ 1\leq i,j\leq r; \\
\{\bar{q}_k, \bar{q}_l\}_D&=&(\delta _{kl}-\frac {1}{n+m}) \delta^{'},  \ \ 1\leq k,l\leq n+m,     \nonumber
\end{eqnarray}
while the three groups of fields $\bar{p_i}$, $\bar{q_k}$ and $J$ mutually commute.
Let
\begin{eqnarray}
\label{e51}
\bar{L}_{1}(p)&=&\prod\limits_{k=1}^{n+m}(p+\bar{q}_k)=
L_{1}(p-\frac{m}{2}J);  \nonumber\\
\bar{\bar{L}_{2}}(p)&=&\prod\limits_{i=1}^{r}(p+\bar{p}_{i})^{m_i}=L_{2}
(p-\frac{m+n}{2}J).
\end{eqnarray}

Thus the Poisson structures associated with $\bar{L}_1$ and $\bar{L}_2$ give rise
to ${\rm w}_{n+m}$ and ${\rm w}_{[m_1,\cdots,m_r]}$ algebras respectively. Togather
with {\rm U(1)} current $J$, they indepent to each other. The
transformation between original fields $(p_i, u_l,a_{ij})$ and new fields
$(\bar{p}_i, \bar{q}_k,J)$ are given by the identity
\be
\label{e52}
L(p)=\bar{L_1}(p+\frac{m}{2}J)\bar{L}_2^{-1}(p+\frac{m+n}{2}J).
\ee
According to Theorem 3, the free fields representation of ${\rm w}(n, [m_1, \cdots , m_r])$
-algebra can be formulated by using the free fields representations of ${\rm w}_{n+m}$ and
${\rm w}_{[m_1, \cdots , m_r]}$ algebras. \par
\begin{coro}
If $r=m, m_i=1$, then ${\rm w}(n,[1,1,\cdots, 1])$=${\rm w}(n,m)$ is nothing
but the classical limit of ${\rm W}(n,m)$-algebra defined in [14], and is
isomorphic to ${\rm w}_{n+m}\bigoplus {\rm w}_m \bigoplus {\rm U}(1)$ via a
transformation. The last result is an analogue of that for ${\rm W}(n,m)$-algebra
[14].
\end{coro}

\par
\begin{rem}
The proof of Theorem 3 is to construct an explicit transformation. similar
transformation can also be constructed for the ${\rm w}(n,m)$-algebra such that the
result of [14] that ${\rm W}(n,m)\cong {\rm W}_{n+m}\bigoplus {\rm W}_{m}
\bigoplus {\rm U(1)}$ can be proved straightforward by this transformation.
\end{rem}
\par
\par
In conclusion, we have discussed the Poisson structures for the dispersionless
systems and associated w-algebra. We are particularly interested in the case
that the $n^{th}$-order polynominal $L$ has multiple roots or the c$\Psi $DO
$L$ has multiple poles at finite distance, and we gave the so-called ${\rm w}_{[m_1,\cdots,
m_r]}$ or ${\rm w}(n, [m_1, \cdots, m_r])$ algebras and their free fields representations. \par
In [7], a dispersionful analogue of the Benney hierarchy associated with
c$\Psi $DO $L$ that has simple poles at finite distance (i.e. $r=1$ and
$m_i=1$ in (42)) was constructed. This dispersionful system can be identified with the
multicomponent generalization of the constrained KP hierarchy [15]. It would
be very interest to see whether there exists dispersionful analogue of the
extended Benney hierarchy associated with $L$ having multiple poles. In
other words, we would ask whether there exists a W-algebra that take our
${\rm w}(n, [m_1, \cdots , m_r])$ as its dispersionless limit. \par
\vskip 0.3in
\section* {Acknowledgement}  \par
This work was supported by the National Basic Research Project for
"Nonlinear Science" and Fund of Education Committee of China. \par

\end{document}